\documentclass[12pt]{article}

\usepackage{graphics}
\usepackage{amsmath}
\def\ni{\noindent}
\def\ph{{\phantom{...}}}
\def\={\phantom{..} = \phantom{..}}

\def\+{\phantom{..} + \phantom{..}}
\def\>{\phantom{..} > \phantom{..}}
\def\<{\phantom{..} < \phantom{..}}
\def\-{\phantom{..} - \phantom{..}}

\def\bq{\begin{quote}}
\def\eq{\end{quote}}
\def\be{\begin{equation}}
\def\ee{\end{equation}}
\def\bar{\begin{eqnarray}}
\def\ear{\end{eqnarray}}
\def\no{\nonumber}

\def\Sch{Schr{\"o}dinger}

\def\Re{\hbox{Re}}

\def\bO{{\hbox{O}}}

\def\sjkN{\sum_{j,k=1}^N}
\def\skN{\sum_{k=1}^N}

\def\Re{\hbox{Re}}

\def\Ntup{x_1,x_2,...,x_N}
\def\ootm{\left(\,\frac{1}{2\,m}\,\right)}  
  
\def\ootri{- \ootm\,\triangle}

\def\ovx{\overline{x}}
\def\pdxk{\prod\,dx_k}

\def\cC{{\cal C}}

\def\smu{\sum_{\mu = 1}^3}
\def\sjkN{\sum_{j,k:j \neq k}}
\def\gsig{g_{\sigma}}
\def\intdthreex{\int\,\prod_{j=1}^N\,dx^3_j}

\def\eU{\epsilon_{U}}

\title{\bf Islands of Instability in Nonlinear Wavefunction Models in the Continuum:\\[0.5in]
A Different Route to ``Chaos"\\[3in]}

\author{W. David Wick\footnote{email: wdavid.wick@gmail.com}}

\begin{document}
\maketitle
\pagebreak

\section*{Abstract}
In two previous papers the author described ``Islands of Instability" that may appear 
in wavefunction models with nonlinear evolution (of a type proposed originally 
in the context of the Measurement Problem). 
Such ``IsoI" represent a new scenario for Hamiltonian systems
implying so-called ``chaos".
Criteria was derived for, and shown to be fulfilled in, 
some finite-dimensional (multi-qubit)
models, and generalized in the second paper to continuum models. But the only example produced
of the latter was a model whose linear \Sch\ equation was exactly-solvable. As
exact solutions of many-body problems are rare, here I show that the instability criteria
can be verified by plugging test-functions into certain computable expressions, bypassing
the solvability blockade. The method can accommodate realistic inter-molecular
potentials and so may be relevant to instabilities in fluids and gasses.

\section{Introduction}

After introducing novel terms into \Sch's equation (in paper I of a series, \cite{WickI})
now said to represent ``WaveFunction Energy" (WFE)
and intended as a solution to the Measurement Problem, 
this author addressed the accompagning Randomness Problem: why do outcomes in certain
experiments appear as if dice had been thrown? He briefly settled on a universal random
component of wavefunctions (paper II; \cite{WickII}), then abandoned the idea after
recalling that high-dimensional, nonlinear dynamical systems often exhibit so-called ``chaos"
(paper III; \cite{WickIII}). In a later paper (\cite{WickChaos}),
he demonstrated Sensitive Dependence on Initial Conditions (SDoIC) for 
a three-qubit model through simulations, using established criteria. It was at that time
that he noticed a novel phenomena: the existence of open regions in which the linear predictor
of the dynamics reported stable and unstable directions meeting at a point,
which might therefore be called ``Islands of Instability" (IsoI). The simulated system entered
and departed such regions repeatedly. Moreover, the appearance of IsoI coincided
with the positivity of a Lyapunov exponent (the formal justification for claiming ``chaos").

IsoI is not the explanation for chaos in Hamiltonian systems preferred by most mathematicians.
Rather, chaos is usually supposed to derive from, or even be identical with, the appearance
of many periodic solutions of the equations with different periods (even a dense family
of such periodic orbits).\footnote{Historically, this preoccupation may
derive from the hope to apply the theory of iterated maps: by looking at a cross-section,
periodic orbits yield a ``return map" that is iterated.
I note however that
the maps in classic examples of ``chaos" were dissipative or discontinuous, hence not Hamiltonian.} 
Thus a small perturbation will knock the system onto a nearby
orbit of different period, which, on some time scale, implies divergence from the original
trajectory. By contrast,
the IsoI conception is of occasional ``crisis" episodes in which that small perturbation
may send the system off at right angles to the current path. The latter seems more relevant
to measurement situations, which by their nature must possess such crises (as every measurement
on a microsystem requires an amplification step).  

Perhaps the distinction between the two conceptions of chaos can be illuminated by
considering the Solar System, thermodynamic systems such as liquids and gases, and the game of 
pinball. For the SS, thirty years ago a team, \cite{MandH},
 simulated the outer planets on a fast computer,
twice, making a 1.5 centimeter displacement of the center-of-mass of Uranus in the second round.
The orbits diverged on a scale of millions of years. The authors explained what is going
on by citing the dense-periodic-orbits scenario. Now, I assume they believed that
this form of SDoIC would appear with {\em any} initial conditions and {\em any} small perturbation
(say, a 1.7 centimeter shift in the position of Saturn). Thus a ``global" SDoIC. 

But does every object currently inhabiting 
the SS have periodic orbits? No! Astronomers routinely sight
comets and asteroids with high velocities and hyperbolic orbits. These extra-solar visitors
swing around the sun and depart in some direction, never to be seen by humans again.
For these objects, a small perturbation in their initial direction of motion may
produce a big variation in that final trajectory---so we might diagnose
a case of SDoIC. But it is clearly not a global statement of chaos.

Next, consider the postulates of thermodynamics---in particular, that the system
over time will visit every part of the configuration space, limited only by constraints.
Clearly, we demand that this behavior be universal and not limited to particular
starting configurations. Does that mean that every configuration must be unstable---or
might it be sufficient if every system trajectory experiences some eras of instability over time?

Finally, consider the game of pinball, in which seemingly random outcomes derive from
complicated trajectories and SDoOC. From considering the initial trajectory,
this is not a case of global chaos; indeed, unless properly aimed, the ball may miss
the first pin and continue on its (stable) way. Thus, I claim pinball is similar to measurement
situations in atomic physics, which necessarily create unstable trajectories if 
anything microscopic is to be measured.

Returning to wavefunctions and nonlinear evolution laws,
in the previous chaos paper the author made an analytic study of the 
what was called there the ``Determinant Criterion for Instability" (DCI).
This study resulted in replacing  conditions on a matrix, which do not readily generalize
to continuum models, by a set of inequalities, which do.
However, these operator inequalities were difficult to check, and as a result the only example
displayed was a collection of bodies in one spatial-dimension in a harmonic potential well,
with no inter-body interactions (an exactly-solved model minus the nonlinear terms).
The goal in this work is to replace the operator inequalities in the criterion
by bounds produced by plugging selected wavefunctions into various expressions
given {\em a priori}, which
obviates having to exactly solve the linear model.

In addition to introducing the analytical conveniences indicated above,
I extend the allowable models to have three spatial dimensions,
as well as realistic inter-molecular potentials,
such the ever-popular Lennard-Jones. Using the new methodology, I am able
to discuss the nature of the instability discovered in the earlier work,
in particular how it depends on system-size and on the essential distinction
between classical and wavefunction models: namely, the latter permit
superpositions unknown to the older physics.

Section \ref{Modelsection} introduces the model class to be considered. Section \ref{ICsection}
reworks the instability criteria. Section \ref{NCsection} utilizes a ``nearly-classical"
wavefunction to plug in and make some observations. Section \ref{Supersection} shows how
superpositions can generate instability. Section \ref{Boundingsection} contains
a rather technical discussion of why we can expect the ``strength of instability"
to pass from discrete to continuous models (details in a Math Appendix). Finally,
in section \ref{Discussionsection} I discuss what the relevance of this work 
may be to realistic scenarios.

\section{The Models, in Three Space Dimensions and with 
Inter-molecular Potentials.\label{Modelsection}}

The criteria are basically the obvious generalizations from paper \cite{WickChaos}.
The arguments of the wavefunction, $\psi$, are now: $x_{k;\mu}$, where $k = 1,2,...,N$
and $\mu = 1,2,3$. I will frequently use $x_k$ for the vector in $R^3$ with components
$x_{k,1}, x_{k,2}, x_{k,3}$.
Define (setting $\hbar = 1$):

\be
\Lambda \=  \ootri \+ V(\Ntup),
\ee

\ni where

\be
\triangle \= \skN\,\smu\,\frac{\partial^2}{\partial x_{k;\mu}^2}.
\ee

\ni For the potential written as $V(\cdot)$ I take:

\bar
\no V(\Ntup) &\=& B(\Ntup) \+ U(\Ntup);\\
\no B(\Ntup) &\=& \left(\,\frac{v}{2}\,\right)\,\skN\,||x_k||^2;\\
\no U(\Ntup) &\=& \sjkN\,u(\,||x_j - x_k||\,).\\
&&
\ear

\ni where `$v$' is a positive parameter. The term written `$B$' represents
a Harmonic well trapping the system bodies and so defining a characteristic
length parameter. 
For the inter-molecular pair-potential, the iconic example is that of Lennard-Jones (LJ):

\be
u(x) \= 4\,{\cal E}\,\left\{\,\left(\,\frac{\alpha}{x}\,\right)^{12} 
\- \left(\,\frac{\alpha}{x}\,\right)^{6}\,\right\},
\ee 

\ni which is graphed in Figure 1. The positive parameters ${\cal E}$ and $\alpha$ 
have units of energy and length, respectively.\footnote{Usually
the length parameter is denoted by $\sigma$ but I need that letter to represent 
a variance later.} In classical statistical mechanics, the Lennard-Jones
potential---due to its repulsion
at the origin, negative global minimum ($- \epsilon$, located at $2^{1/6}\,\alpha$),
and rapid fall-off at infinity---has been shown to support 
solid, liquid, and gas phases and even a triple-point in the phase diagram.
Hence it's popularity with physicists and chemists.  

\begin{figure}
\rotatebox{0}{\resizebox{5in}{5in}{\includegraphics{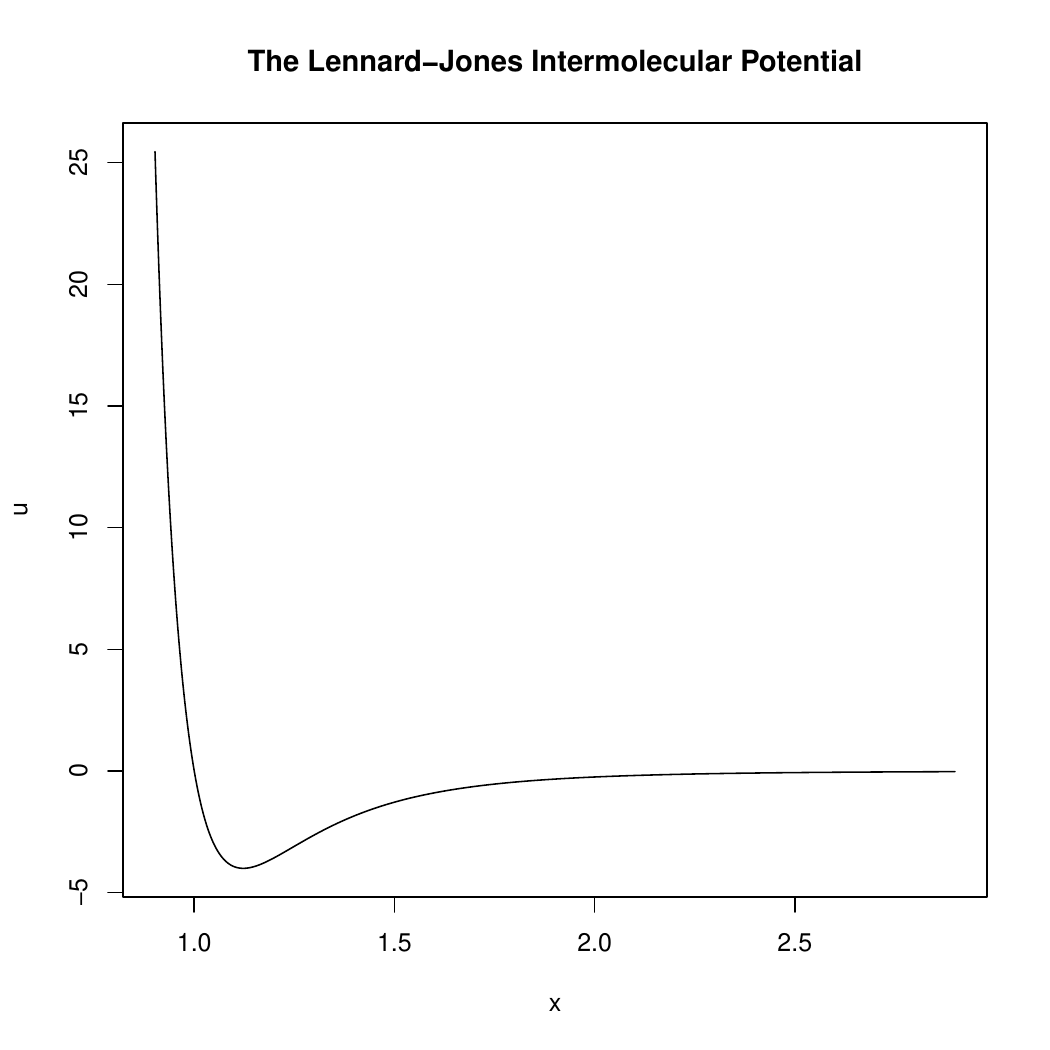}}}
\caption{The Lennard-Jones potential, with ${\cal E } = 4.0$,
$\alpha = 1.0$. }\label{Fig1}
\end{figure}

A key property of the L-J potential is ``(thermodynamic) stability",
meaning that their exists a positive parameter, call it $\epsilon_U$, such that:

\be
U(\Ntup)  \geq - e_U\,N\label{stabilitycond}
\ee 

\ni holding for all $\Ntup$,  
see, e.g., \cite{Gall}, \cite{Pro}.
Mathematicians working in classical statistical mechanics have abstracted the 
form to define a class of potentials of ``Lennard-Jones type":

\bar
 \no u(x) &\geq& \frac{c_1}{x^{3 + \epsilon}},\ph \hbox{if}\ph x \leq a;\\
\no |u(x)| &\leq& \frac{c_2}{x^{3 + \epsilon}}, \ph \hbox{if}\ph x \geq a,\\
&&\label{LJcond}
\ear

\ni for positive parameters $c_1,c_2$. These properties of the pair-potential
somewhat complicate the domain question for the operator $\Lambda$, as the
contribution to the total energy:

\be
\intdthreex\,|\psi|^2\,U(\Ntup),
\ee

\ni may be infinite without requiring $\psi$ to be zero (and drop to zero rapidly)
on the coincidence set: $\cC: \{\,x_j = x_k\ph \hbox{for some}\, j \neq k\,\}$.
This is because a potential of ``L-J type" is not locally integrable on $R^3$.

However, the conditions in (\ref{LJcond}) are far from necessary. 
Procacci gives the example (\cite{Pro}, p. 45):

\be
u(x) =
\begin{cases}
 A, & \hbox{if}\ph x \leq R\\
 -1, & \hbox{if}\ph R < x \leq R + \delta\\
0,& \hbox{otherwise}.
\end{cases}\label{Bas}
\ee

\ni which he demonstrates is stable---satisfies (\ref{stabilitycond})---if $A > 12$
and $\delta$ is small enough. Note that, although discontinuous, this potential
is bounded above, so that no such restrictions on the domain of $\Lambda$ would be
necessary. Also, clearly one could draw a smooth potential approximating
this example and hence stable.

The potential in (\ref{Bas}) was chosen as an instance of a general criterion due to
Basuev (\cite{Pro}, p. 43) which evidently provides a large class of
stable potentials bounded above and with integrable tails (such potentials are called ``tempered"). 
I will refer to this class 
as constituting the ``integrable potentials".

\section{The Instability Criteria, in three space dimensions.\label{ICsection}}

We will need these quantities:

\bar
\no f &\=& w\,\ovx^t\,\left(\,\ovx - 2\,S\,\right);\\
\no \ovx &\=& \skN\,x_k;\\
\no S &\=& \int\,\prod\, d\,x_k\,|\psi(\Ntup)|^2\,\ovx;\\
&&
\ear

Here the superscript `$t$' means transpose and $\ovx$ and $S$ are 3-vectors.
The positive parameter `$w$' represents the coupling constant between the conventional 
energy and
the WFE; it's order-of-magnitude is currently unknown.

The nonlinear dynamics is then generated from the Hamiltonian:

\be
H \= <\psi\,|\Lambda|\,\psi> \+ w\,\left\{\,<\psi\,|\,||\ovx||^2\,|\,\psi> \- ||S||^2\,\right\}.
\ee

See \cite{WickI} for Weinberg's simple derivation of a nonlinear \Sch's equation
from this Hamiltonian,
or \cite{WickLagrange} for a Lagrangian approach.

The criteria from the previous chaos paper then read:

\begin{quote}

(i) 
\be
\Lambda > 0;
\ee
(ii) 
\be
\Lambda \+ f(\Ntup) > \eta > 0;
\ee
(iii) 
\be 
4\,w\,\int\,\prod\,dx_k\,\left[\,\hbox{Im}\,\psi(\Ntup)\,\right]^2 \,\ovx^2 < \eta;
\ee
(iv) 
\be
f(\Ntup) \leq \left(\,\rho - 1\,\right)\,\,\Lambda,
\ee
\ni for some $\rho > 1$;\footnote{In the previous paper the character used 
here was $\sigma$ but I will need that symbol later for a variance.}

(v) 
\be
4\,w\,\int\,\pdxk\,\left[\,\Re\,\psi\,\right]\,\ovx^t\,\Lambda^{-1}
\,\ovx\,\left[\,\Re\,\psi\,\right]\, \geq\, \rho.\label{fifthcondition}
\ee

\end{quote}

In (i), (ii), and (iv), the inequalities are meant as between operators on
some Hilbert space of functions (which can be thought of as the tangent space at $\psi$); hence,
both $\psi$ and $S$, which appears in $f$, should be regarded as being constants.

The proof of (i) in our class of models procedes by noting that

\be
\ootri \+
\left(\,\frac{v}{2}\,\right)\,\skN\,||x_k||^2 \geq \omega\,\left(\,\frac{3\,N}{2}\,\right),
\ee

\ni where 

\be
\omega \= \sqrt{\,\frac{v}{m}\,}.
\ee

\ni Hence

\be
\Lambda \geq 
\omega\,\left(\,\frac{3\,N}{2}\,\right) \- \eU\,N,
\ee

\ni which is positive provided 

\be
\frac{3\,\omega}{2} > \eU.
\ee

The proof of (ii) then procedes as in the previous paper.

For (iii) take $\hbox{Im}\,\psi = 0$.

For the proof (iv) we rewite it as:

\be
f \leq \left(\,\rho - 1\,\right)\,\left[\,- \,\left(\,\frac{1}{4m}\,\right)\,\triangle\,
+ \frac{B}{2} 
- \,\left(\,\frac{1}{4m}\,\right)\,\triangle\,
+ \frac{B}{2} \, +\, U\,\right].
\ee

\ni Now choose $v,m$ so that

\be
- \,\left(\,\frac{1}{4m}\,\right)\,\triangle\,
+ \frac{B}{2} \, +\, U \geq 0.
\ee

Hence it suffices to prove

\be
f \leq \left(\,\rho - 1\,\right)\,\left[\,- \,\left(\,\frac{1}{4m}\,\right)\,\triangle\,
+ \frac{B}{2}\,\right]. 
\ee

\ni Now the proof of (iv) in the previous paper can be followed, but with
$v$ replaced by $v/2$, $m$ by $2\,m$, and $N$ by $3N$.

Hence the principal issue, which will be the focus of this paper, is to find
wavefunctions that, introduced into part (v), give the least restriction on `$w$'.
Defining:

\be
\Omega \= \ovx^t\,\Lambda^{-1}
\,\ovx \= \smu\,\ovx_{\mu}\,\Lambda^{-1}\,\ovx_{\mu} \= \smu\,\Omega_{\mu}.
\ee

\ni we are interested in the upper bound on it, in other words

\be
\sup\,\left\{\,\frac{<\psi\,|\,\Omega\,\psi>}{||\psi||^2}\,\right\},
\ee

\ni which will of course occur at an eigenfunction:

\be
\Omega\,\psi \= \lambda\,\psi.
\ee

\def\lmax{\lambda_{\hbox{max.}}}

Let the largest such eigenvalue be $\lmax$.\footnote{Of course, $\lmax$ might be infinite,
but then condition (v) is vacuous.}
Noting that $\Omega$ is a sum of positive operators, we have

\be
\lmax \geq \max_{\mu = 1,2,3}\,\sup_{\psi}\,
\left\{\,\frac{<\psi\,|\,\Omega_{\mu}\,\psi>}{||\psi||^2}\,\right\},\label{maxsup}
\ee

\ni hence we can get a lower bound on $\lmax$ 
by substituting a wavefunction into each expression separately.

\def\Omu{\Omega_{\mu}}
\def\Omuisr{\Omega_{\mu}^{-1/2}}
\def\Omumo{\Omega_{\mu}^{-1}}

Noting that $\Omu$ is a positive operator, its inverse square-root, $\Omu^{-1/2}$,
is a well-defined, bounded (hence continuous) operator. 
It will be useful to plug into (\ref{maxsup}) functions of
the form: $\psi = \Omuisr\phi$ transforming
the problem to:

\bar
\no \lmax &\geq& \sup\,\left\{\,\frac{<\Omuisr\,\phi\,|\,\Omu\,\Omuisr\,\psi>}
{<\Omuisr\,\phi|\Omuisr\,\phi>}\,\right\}\\
\no  &\geq& \sup\,\left\{\,\frac{||\phi||^2}{<\phi\,|\,\Omumo\,\phi>}\,\right\}\\
\no  &\geq& \left\{ \inf_{\phi}\,\left[\,\frac{<\phi\,|\,\Omumo\,\phi>}{||\phi||^2}\,
\right]\,\right\}^{-1}\\
&&
\ear

\ni Thus, plugging into

\be
\frac{<\phi\,|\,\Omumo\,\phi>}{||\phi||^2}\label{Oratio}
\ee

\ni will generate a lower bound on $\lmax$. Note that:

\def\ovxmumo{\overline{x}_{\mu}^{-1}}
\def\ovxmu{\overline{x}_{\mu}}

\be
 \Omumo \=  \ovxmumo \,\Lambda\,\ovxmumo,
\ee

\ni which is considerably easier to work with than $\Omu$.

A simple calculation yields, for the numerator in (\ref{Oratio}):

\bar
\no && - \left(\,\frac{1}{2m}\,\right)\,\left\{\,2\,N\,<\phi|(\ovxmu)^{-4}\,\phi> -2\,\skN\,
<\phi|(\ovxmu)^{-3}\,\frac{\partial\,\phi}{\partial \, x_{k,\mu}}> + \right.\\
\no && \left. \skN\,\sum_{\nu = 1}^3\,<\phi\,|(\ovxmu)^{-2}\,\frac{\partial^2\,\phi}
{\partial\,x_{k,\nu}^2}>\,\right\} \+
<\phi\,|\,(\ovxmu)^{-2}\,V\,\phi>.\\
&&
\ear

Now make another substitution: $\phi = \ovxmu^2\,\theta$ to get:

\bar
\no && - \left(\,\frac{1}{2m}\,\right)\,\left\{\,2\,N\,||\theta||^2 -2\,\skN\,
<\theta|(\ovxmu)^{-1}\,\frac{\partial\,(\,\ovxmu^2\,\theta)}{\partial \, x_{k,\mu}}> + \right.\\
\no && \left. \skN\,\sum_{\nu = 1}^3\,<\theta\,|\,\frac{\partial^2\,(\,\ovxmu^2\,\theta)}
{\partial\,x_{k,\nu}^2}>\,\right\} \+
<\theta\,|\,\ovxmu^{2}\,V\,\theta>.\\
&&
\ear

Applying Leibniz's rule in the second term and performing an IBPs (assuming
that $\theta$ vanishes rapidly together with all its derivatives at infinity) on the third yields:

\bar
\no && - \left(\,\frac{1}{2m}\,\right)\,\left\{\,2\,N\,||\theta||^2
 - 4\,\skN\,<\theta|\theta>  -2\,\skN\,
<\theta|\ovxmu\,\frac{\partial\,\theta}{\partial \, x_{k,\mu}}> \- \right.\\
\no && \left. \skN\,\sum_{\nu = 1}^3\,<\frac{\partial \theta}{\partial \,x_{k,\nu}}\,|\,
\frac{\partial\,(\,\ovxmu^2\,\theta)}
{\partial\,x_{k,\nu}}>\,\right\} \+
<\theta\,|\,\ovxmu^{2}\,V\,\theta>.\\
&&
\ear

\ni With another use of Leibniz on the last term in curly brackets this simplifies to:

\bar
\no && \left(\,\frac{1}{2m}\,\right)\,\left\{\,2\,N\,||\theta||^2
 + 4\,\skN\,\hbox{Re}\,<\theta|\ovxmu\,\frac{\partial\,\theta}{\partial \, x_{k,\mu}}> + \right.\\
\no && \left. \skN\,\intdthreex\,\ovxmu^2\,||\frac{\partial \theta}{\partial \,x_k}||^2\,\right\}
 \+ <\theta\,|\,\ovxmu^{2}\,V\,\theta>.\\
&&
\ear

Note that, using the same tricks,

\bar
\no && \intdthreex\,\theta^*\,\ovxmu\,\left[\,\frac{\partial\, \theta}
{\partial \,x_{k,\mu}}\,\right]
\+ \intdthreex\,\theta\,\ovxmu\,\left[\,\frac{\partial\, \theta^*}{\partial \,x_{k,\mu}}\,\right]\\
\no &\=& \intdthreex\,\ovxmu\,\left[\,\frac{\partial\,( \theta^*\,
\theta)}{\partial \,x_{k,\mu}}\,\right]\\
\no &\=& \- \intdthreex\,|\theta|^2\\
\no &\=& \- \,||\theta||^2,\\
&&
\ear

\ni so that the numerator is simply:

\be
 \left(\,\frac{1}{2m}\,\right)\,\left\{\,
 \skN\,\intdthreex\,\ovxmu^2\,||\frac{\partial \theta}{\partial \,x_k}||^2\,\right\}
 \+ <\theta\,|\,\ovxmu^{2}\,V\,\theta>.
\ee

The denominator is

\be
\intdthreex\,\ovxmu^4\,|\theta|^2.
\ee

In conclusion, we have the lower bound:

\be
\lmax \,\geq\,\left[\,
 \frac{\left(\,1/2m\,\right)\,
 \skN\,\intdthreex\,\ovxmu^2\,||\partial \theta/\partial \,x_k||^2
 + <\theta\,|\,\ovxmu^{2}\,V\,\theta>}
{\intdthreex\,\ovxmu^4\,|\theta|^2}\,\right]^{-1}.\label{lbeq}
\ee

\ni Thus, we wish to plug in a function $\theta(x_1,x_2,...,x_N)$ 
that makes the ratio inside the square brackets
as small as possible, pushing $\lmax$ as high as possible.

\section{A ``Nearly Classical" Test Wavefunction.\label{NCsection}}

From the last section we have the goal of finding a test function 
$\theta(x_1,x_2,...,x_N)$ making the expression: 

 \be
\left(\,1/2m\,\right)\,
 \skN\,\intdthreex\,\ovxmu^2\,||\partial \theta/\partial \,x_k||^2
 + <\theta\,|\,\ovxmu^{2}\,V\,\theta>\label{numexp}
\ee

\ni as small as possible, while making the expression:

\be
\intdthreex\,\ovxmu^4\,|\theta|^2\label{denomexp}
\ee

\ni as large as possible.

With the inter-molecular potentials of ``integrable" class and negative tails,
  it is of immediate interest
to try and maximize the tail-contribution from $U$ in the numerator. 
One candidate might be constructed from the Gaussian:

\be
\gsig(x) \= \left(\,\frac{1}{\sqrt{2\,\pi}\,\sigma}\,\right)^{3/2}\,\exp\left\{\,
- ||x||^2/(4\,\sigma^2)\,\right\}.
\ee

\ni by adopting a regular lattice $\{y_k\}$, $y_k \in R^3$, 
and defining:

\be
\theta(x_1,...,x_N) \= \prod_{k=1}^N\,g(x_k - y_k),
\ee

Trying for a minimum energy, let the lattice spacing be `$a$' and 
set $a $ equal to the location of the global minimum of $U$, with $U(a) = - \epsilon$.
Then from nearest neighbor pairs with $||y_j - y_k|| = a$ there will
a contribution to U of approximately $- \epsilon$ 
from configurations with $||x_j - x_k|| \approx ||y_j - y_k||$.
If $\sigma$ is smaller than $a$, there will be a smaller contribution from
pairs with $||y_j - y_k|| = 2\,a$, and so forth. 
Due to the integrable fall-off of the potential and the super-exponential decay of the Gaussian,
we can expect a convergent series and a contribution from the second term 
like $- \eU\,N$. Since, for part (i), we have assumed that $B$ dominates $U$,
and $B$ also contributes $\bO(N)$, we cannot impose on the second term in the numerator
that it be negative. 

What is the contribution from the first term in the numerator?
From the standard formula for a Gaussian integral we obtain:

\bar
\intdthreex\,|\theta|^2 &\=& 1;\\
 \intdthreex\,|\frac{\partial\,\theta}{\partial x_{k,\mu}}|^2 &\=& 
\frac{1}{16\,\sigma^2}.\label{gaussianintegrals}
\ear

For the contribution of factors containing powers of $\ovxmu$, an easy computation
gives:

\be
\intdthreex\,|\theta|^2\,\ovxmu^2 \= \sum_k\,\sum_j\,y_{j,\mu}\,y_{k,\mu} \+ N\,\sigma^2.
\ee

\ni By centering the lattice we can make the first term vanish. 
Hence $\ovxmu = \bO(\sqrt{N}\,\sigma)$. 

For this ``nearly-classical" state all terms in the ratio in (\ref{lbeq})
are $\bO(N^2)$. Thus we cannot rely on a ``large N" scenario to yield an example
fulfilling the instability criteria, although they may hold for some ranges of 
parameters.

\section{``Large N" Scenario and the Role of Superpositions.\label{Supersection}}

It is the possibility of forming superpositions which is the essence of
wavefunction physics, differentiating it from classical physics (as well
as the source of the Measurement Problem). So let us consider superpositions
of spatially-translated states of the type presented in the last section. I.e.,
states of form

\be
\left(\,\frac{1}{\sqrt{2}}\,\right)\,\theta_{-R} \+ 
\left(\,\frac{\beta}{\sqrt{2}}\,\right)\,\theta_R,
\ee

\ni where $\beta$ is a complex number of modulus one and:

\be
\theta_{\pm R}(x_1,x_2,...,x_N) \=   
\theta(x_1 \pm R\,e_{\mu},x_2 \pm R\,e_{\mu},...,x_N \pm R\,e_{\mu}).
\ee

\ni where $e_{\mu}$ is a unit vector in the $\mu$-direction.

For either translated state, if `$R$' is large enough $\ovxmu \sim \bO(N)$.
Now the estimated magnitudes of the terms in  (\ref{lbeq}) change: the two
terms in the numerator become $\bO(N^3)$ while the term in the denominator 
becomes $\bO(N^4)$. 

Because the upper bound required to fulfill criterion (iv) was $\bO(1/N)$,
we can now see a case arising fulfilling all the criteria (by leaving a gap
between lower and upper bounds in which `$w$' could reside).

\section{Bounding the Positive Eigenvalue as 
Discrete Approaches Continuous.\label{Boundingsection}}

What might we call the ``strength" of the instability? In the 
finite-dimensional (qubit) models, this notion is embodied in 
the positive eigenvalue of the Jacobian matrix written `$M$' in previous papers
(and presumably related to the Lyapunov exponent).
However, the matrix
set-up has no obvious extension to continuum models. 

A plausible resolution of this conundrum is to start with a discrete model,
say by replacing space by a three-dimensional lattice of points, and pass 
in a limit to the continuum model. If the lowest positive eigenvalue of $M$
stays bounded below (away from zero), we can conclude that the instability
persists in the latter case (for, if space is actually discrete at some
extremely small scale, as some physicists have proposed, we could simply 
regard the continuum as a convenient fiction, as we do for the thermodynamic
limit). 

\def\Ne{N_{\epsilon}}
\def\Fe{F_{\epsilon}}
\def\Ge{G_{\epsilon}}
\def\lame{\lambda_{\epsilon}}

Accordingly, let the lattice spacing be $\epsilon$. If we imagine that the system
is contained in a box of side $`L$', then we will need $(L/\epsilon + 1)^3 = n$
lattice points, and we will be interested in $\epsilon \to 0$ and $n \to \infty$.
With `$N$' bodies ($N$ fixed), the dimension of the (real) Hilbert space 
is then $N_{\epsilon} = 2\,n^N$.
For suitably smooth functions, we can derive an $\Ne$-dimensional vector by evaluating
at lattice points. Ditto for the potential-energy functions.
For the kinetic energy operator, $(-1/2\,m)\,\triangle$,
we replace the Laplacian by a second-order, finite-difference operator, $\triangle_{\epsilon}$, 
using $\epsilon$
as the differencing length. 

In the Math Appendix I show that the characteristic polynomial of $M$
can be written:

\be
\det\,\left(\,M - \lambda\,I\,\right) \= F_{\epsilon}(\lambda)\,\cdot\,G_{\epsilon}(\lambda),
\ee

\ni with $\Fe(\lambda) \neq 0$ on some interval $[0,\lambda^*]$ although
it may be the case that $\Fe(\lambda) \to \infty$ or zero as $\epsilon \to 0$,
while $\Ge(\lambda)$ has a limit as $\epsilon \to 0$. Suppose that the first zero
of $\Ge(\lambda)$ to the right of the origin and in the interval:

\be
\lame \= \inf\,\left\{\, 0 < \lambda < \lambda^*:\ph \Ge(\lambda) = 0\,\right\},
\ee

\ni enjoys the bound

\be
\lame \geq \lambda_{+} > 0,
\ee

\ni as $\epsilon \to 0$, which should follow if $G_{\epsilon}$ has a limit. 
This is the desired conclusion.

Perusing the Math Appendix, it will become clear that the form of convergence,
discrete to continuous, we need might be labelled ``resolvent" or ``inverse-operator"
convergence. That is, we need that expressions like

\be
v_{\epsilon}^t\,R_{\epsilon}^{-1}\,u_{\epsilon}
\ee

\ni converge as $\epsilon \to 0$, where $u_{\epsilon}$ and $v_{\epsilon}$ are vectors 
obtained by evaluating (suitably-smooth) functions on lattice sites,
and $R_{\epsilon}$ is an invertible operator on the finite-dimensional Hilbert space obtained
in a natural way from the operators on the continuum Hilbert space defining the problem.

\section{Discussion.\label{Discussionsection}}

Admittedly, the ``nearly-classical" state introduced in section \ref{NCsection}
seems more like the wavefunction
of a crystal rather than that of a liquid or gas. 

Concerning the superposition state considered in section \ref{Supersection},
if $N$ is too large it might constitute a `cat'.
The goal of introducing nonstandard terms into the Hamiltonian was to
eliminate just such an occurrence, assuming the system has insufficient energy
to form it. Thus we are confronted with a competition: a system that
our criteria shows to be unstable might also be cat-like, so will not
be possible given an energy limitation. Alternatively, the instability may form
at some lower value of `$N$' that would not be thought of as `macroscopic'
 and still be relevant, as it gets amplified during
a subsequent measurement.

Thus we detect a coincidence between forming 
spatial superpositions and proving the instability criteria (albeit in a single,
and rather too ``classical", instance). 
This (weakly) suggests that the Infamous Boundary
may represent also a stability boundary, with the nonclassical (wavefunction)
side exhibiting both phenomena, and the classical side, neither.\footnote{Of course,
there is ``classical chaos", but presumably that appears on a much different
time-scale then what we discuss here.} 

Concerning the discrete-to-continuous limit discussed in section \ref{Boundingsection}: 
there is a large literature developing ``discrete quantum mechanics" or
``discrete \Sch's equation" but the motivation of these authors differs from
mine here. Rather, for the most part they intend either to restore Heisenberg's
original picture of the microscopic world (as occupied by particles 
undergoing jumps\footnote{E.g., see his ``uncertainty paper" of 1927, \cite{Heisenberg},
in which he actually drew an illustration of a ``quantum orbit" by replacing
a ``classical (continuous) orbit" by a series of dots. Of course, when
Bohr heard about this he berated his colleague, causing him (Heisenberg) to publish
an addendum containing a (perhaps insincere, and only partial) 
retraction of his conception.}), 
or hope to discover some new predictions. By contrast, I employ discretization merely
as a technical device to aid proofs (and believe that space is a continuum). 

Finally, concerning ``instabilities in fluids and gasses": the most famous such discovered 
was of course
Reynolds' and other's observations in the late 19th Century 
of an instability in fluid flow down a pipe (which, at certain ``Reynolds numbers"
can exhibit the phenomenon called ``turbulence"). Bulk fluids certainly are on the ``classical"
side of the Infamous Boundary, by which I mean that superpositions of such a large
sample are presumably ruled out by WFE. Nevertheless, small packets of fluid
might exhibit an instability as described here; but for it to propagate to
an extent that it would be noticed by a casual observer would require
some kind of amplification mechanism (or at least neutral stability) existing in its surroundings. 
Another, related, issue is the production of viscosity. If either
phenomena depend on the wavefunction description it would mean that Planck's
constant, as well as the parameter I wrote as `$w$', must appear in the resulting
formulas. But this is a subject for future research.

\section*{Math Appendix\label{mathsection}}

\def\xin{\xi^{(n)}}
\def\etan{\eta^{(n)}}
\def\xione{\xi^{(1)}}
\def\xitwo{\xi^{(2)}}
\def\etaone{\eta^{(1)}}
\def\etatwo{\eta^{(2)}}
\def\xionet{\xi^{(1) t}}
\def\xitwot{\xi^{(2) t}}
\def\etaonet{\eta^{(1) t}}
\def\etatwot{\eta^{(2) t}}

We will need the following theorem from matrix algebra:

\begin{quote}
{\bf Determinant Theorem} For invertible-plus-rank-two (IPR2) matrices:

Let $R$ be an $N\times N$ invertible matrix and $\xin$, $\etan$,
$ n = 1,2$, be row N-vectors. Then:

\bar
\no && \det\,\left(\,R + \xione\,\otimes\,\etaone   
 + \xitwo\,\otimes\,\etatwo \,\right) \=\\
\no && \det R\,\cdot\,\left\{\, \left(\,1 + \etaonet\,R^{-1}\,\xione\,\right)\cdot  
\left(\,1 + \etatwot\,R^{-1}\,\xitwo\,\right)\right.\\  
\no &&\left. \- \etaonet\,R^{-1}\,\xitwo\cdot
 \etatwot\,R^{-1}\,\xione\,\right\}.\\
&&
\ear

{\bf Corollary} 

{\bf Determinant Theorem} For invertible-plus-rank-one (IPR1) matrices:

Let $R$ be an $N\times N$ invertible matrix and $\xi$, $\eta$,
be row N-vectors. Then:

\bar
\no && \det\,\left(\,R + \xi\,\otimes\,\eta   
\,\right) \=\\
\no && \det R\,\cdot\,\left\{\, 1 + \eta^t\,R^{-1}\,\xi\,\right\}.\\ 
&&
\ear

\end{quote}

{\bf Proof of the Theorem}. We make use of: 

\begin{quote}
{\bf Sylvester's Theorem}:\footnote{For Sylvester's Theorem see the Wikipedia page ``Determinants"
or any linear algebra text.}

Let $P$, $Q$ be $N\times n$ and $n \times N$ matrices, respectively. Then:

\be
\det\,\left\{\,I_N + P\,Q\,\right\} \= \det\,\left\{\,I_n + Q\,P\,\right\}.
\ee

\end{quote}

We apply Sylvester's Theorem by first factoring out $R$ and then defining the
matrices:

\be
Q\= \begin{pmatrix}
\etaone_1,&\etaone_2,...& \etaone_N \\
\etatwo_1,&\etatwo_2,...& \etatwo_N \\
\end{pmatrix}
\ee

\be 
P \= \begin{pmatrix}
[R^{-1}\,\xione]_1, & [R^{-1}\,\xitwo]_1 \\
[R^{-1}\,\xione]_2, & [R^{-1}\,\xitwo]_2 \\
. & . & \\
. & . & \\
[R^{-1}\,\xione]_N, & [R^{-1}\,\xitwo]_N \\
\end{pmatrix}.
\ee

It is easily seen that:

\be
I_N + P\,Q \= I_N + R^{-1}\,\xione \otimes \etaone + R^{-1}\,\xitwo \otimes \etatwo,
\ee

\ni and

\be
I_2 + QP \= \begin{pmatrix}
1 + \etaonet\,R^{-1}\xione, & \etaonet\,R^{-1}\xitwo\\
\etatwot\,R^{-1}\xione & 1 + \etatwot\,R^{-1}\xitwo \\
\end{pmatrix}
\ee 

The theorem then follows from the rule for computing determinants of 2x2 matrices. QED
 
{\bf Proof of the Corollary} Set $\etatwo = \xitwo = 0$, $\eta = \etaone$ and
$\xi = \xione$ and apply the result for IPR2. QED

Remark: obviously the theorem can be generalized to any sum: invertible plus
rank-n matrix, yielding the same first factor and some more complicated
sum of terms involving inner products with $R^{-1}$ resulting from computing
the determinant of an $n\times n$ matrix.

Now to apply the above machinery to the characteristic polynomial:

\be
\det\left(\,M - \lambda\,I\,\right) \= \begin{pmatrix}
A - \lambda\,I, & B\\
C, & D - \lambda\,I\\ \end{pmatrix}
\ee

From the basic formula used in \cite{WickChaos}, which did not
depend on the rank of the matrices $A$ and $D$ but only on that $B$ is invertible:

\be
\det\left(\,M - \lambda\,I\,\right) \= 
(-1)^N\,\det B\,\det\left(\,C - [D - \lambda\,I]\,B^{-1}\,[A - \lambda I]\,\right).\label{basiceq}
\ee

For the first factor recall that

\bar
\no B &\=& E - u\,\otimes\,u;\\
\no E &\=& K + \hbox{diag}(f),\\
&&
\ear

\ni where $K$ includes all the contributions from the linear part of the evolution equation,
and $f$ is a certain function derived from the WFE term. Assuming that $E$ is positive
(bounded below) it is invertible and so $\det B$ falls in our competance.  

Defining

\be
R \= C - \lambda^2\,B^{-1} \= B^{-1}\,\left(\,B\,C - \lambda^2\,I\,\right),
\ee

\ni and recalling that $A$ and $D$ are rank-one, we can write the second factor as

\be
\det\left(\,C - [D - \lambda\,I]\,B^{-1}\,[A - \lambda I]\,\right) \= \det\left(\,R + 
\hbox{rank-2}\,\right),
\ee

\ni where the rank-2 matrix can be written down explicitly (involving $B^{-1}$ and
various inner-products). 

Concerning $R$, using 

\be
C = - E + v\otimes v,
\ee

\ni it can be written:

\be
R \= - B^{-1}\,\left(\,E^2 + \lambda^2\,I\,\right),
\ee

\ni and hence is invertible for every $\lambda > 0$ given that $B$ is. Since we have
assumed that $E > 0$, it is still invertible at $\lambda = 0$.

The overall conclusion is that the characteristic polynomial is a product
of factors, some containing determinants of invertible matrices and others
which are polynomials containing terms which are of the form

\be
u^t\,\otimes B^{-1}\,v,
\ee

\ni or

\be
u^t B^{-1}\,R^{-1}\,v.
\ee

Hence the remark in section \ref{Boundingsection} about ``resolvent" or ``inverse operator" convergence.

\section*{Acknowledgement} The author thanks L. DeCarlo for discussion and citations.

\end{document}